\documentclass{aa}

\usepackage{graphics}

\newcommand{\gta}{\mathrel{\hbox{\rlap{\lower.55ex \hbox {$\sim$}} 
                   \kern-.3em \raise.4ex \hbox{$>$}}}} 
\newcommand{\lta}{\mathrel{\hbox{\rlap{\lower.55ex \hbox {$\sim$}} 
                   \kern-.3em \raise.4ex \hbox{$<$}}}} 
 
\begin{document}

   \thesaurus{06     
              (08.02.1;  %
               02.01.2;  %
               02.09.1;  %
               13.25.3;  %
               19.63.1)} %
   \title{X-rays from quiescent low-mass X-ray binary transients}
\titlerunning{X-rays from quiescent transients}

   \author{J.-P. Lasota}

   \offprints{J.-P. Lasota}

   \institute{Institut d'Astrophysique de Paris,
              98bis Boulevard Arago, 75014 Paris, France\\
              email: lasota@iap.fr
             }

  \date{Received; accepted}

   \maketitle

   \begin{abstract}
I argue that it is very unlikely that X-rays from
quiescent black-hole low-mass X-ray binary transients are emitted by coronae
of companion stars. I show that in a simple model in which these X-rays are
emitted by an ADAF filling the inner part of an unsteady, dwarf-nova type
disc, the X-ray luminosity is correlated with the orbital period. I 
predict what values of X-ray luminosities from black-hole transient systems 
should be observed by {\sl Chandra} and {\sl XMM-Newton}.
   \end{abstract}

%

\section{Introduction}

Low-mass X-ray binary transient systems (LMXBTs; these systems are also
called `X-ray novae' or `Soft X-ray transients') are low-mass X-ray
binaries (LMXBs) which sometimes (rarely for the most part) undergo
outbursts during which the X-ray luminosity increases by more than 5 to
6 orders of magnitude. In LMXBs, a black hole or neutron star primary
accretes matter lost by a Roche-lobe filling, low-mass secondary star.
All known low-mass X-ray binaries (LMXBs) containing black holes are
transient (Tanaka \& Shibazaki 1996), whereas many neutron star LMXBs
are steady, in the sense that unlike black-hole LMXBs they do not show
high amplitude outbursts, but only low amplitude X-ray flux variation.
Matter transferred from the secondary, forms an accretion disc, which
far away from the accreting object is quasi-Keplerian. Accretion discs
in LMXBTs appear to be truncated in their inner regions (Esin et
al.1997; \.Zycki et al. 1998,1999). Since here magnetic fields can play
no role (because the magnetic moments of neutron stars are too low, and
because of the absence of black holes magnetic fields) the inner disc
`hole' can be due only to some kind of evaporation (Narayan \& Yi 1995;
Honma 1996). In such a case the inner accretion flow onto the compact
object may form an advection-dominated accretion flow (ADAF; Abramowicz
et al. 1995; Narayan \& Yi 1995). Truncated discs are also required by
the disc instability model (DIM), which is supposed to describe LMXBT
outbursts (Lasota 1996; Menou et al. 2000; Dubus, Lasota \& Hameury
2000). This model was devised to describe dwarf nova (DN) outbursts (see
Cannizzo 1993 and Lasota 2000b for reviews). Also in these systems
truncated discs are required to reconcile models with observations (e.g.
Lasota, Kuulkers \& Charles 1999; Meyer \& Meyer-Hofmeister 1994;
Shaviv, Wickramasinghe \& Wehrse 1999)


According to models of such truncated discs the inner accretion flow is
an optically thin, very hot plasma, in which temperature may be close to
the virial temperature. It is therefore expected to emit a considerable
part of its energy in X-rays. This is indeed observed in quiescent DN
and LMXBTs where such inner hot flows should be present (Eracleus,
Helfand \& Patterson 1991; Richards 1996; van Teeseling, Beuermann \&
Verbunt 1996; McClintock, Horne \& Remillard 1995; Verbunt et al. 1994;
Wagner et al. 1994; Asai et al. 1998; Barret, McClintock \& Grindlay
1996). The properties of this X-ray emission provide an important test of
accretion flow models (see e.g. Quataert \& Narayan 1999; Meyer,
Meyer-Hofmeister \& Liu 1996).

However, since observed X-ray luminosities are often rather low, one
should be sure that the X-rays are not emitted by other sources, in
principle less powerful than accretion. For DN it was shown that
quiescent X-rays are emitted by the accretion flow and not by the
secondaries coronae (Eracleus et al. 1991 van Teeseling et al. 1996;
Richards 1996). For LMXBTs Verbunt (1996) concluded that (``except maybe
for A0620-00") X-rays cannot be emitted by coronae of secondary stars.
In the case of neutron-star LMXBTs Brown, Bildsten \& Rutledge (1998)
attribute the quiescent X-rays to thermal emission from the neutron-star
surface. This emission would be due to repeated deposition during the
outbursts of nuclear energy deep in the crust. This could be a viable
alternative to the accretion model (Rutledge et al. 1999, see however
Menou et al. 1999c).

Recently Bildsten \& Rutledge (1999) concluded that in the case of
black-hole LMXTBs the quiescent X-rays may be due to coronal emission from
stellar companions. They argue that in these systems the ratio of the
X-ray flux to the stellar, bolometric flux is $\lta 10^{-3}$ as in RS
CVn's, which are active, close, detached binaries of late-type stars (a
G of K type giant or subgiant in orbit with a late-type main-sequence or
subgiant) in which, for orbital periods $\lta 30$ days, the rotation of
both components is synchronous with the orbit. Their coronal X-ray
emission may be as large as $10^{31}$ erg s$^{-1}$ (Dempsey et al.
1993). 

Unfortunately, only three quiescent black-hole LMXBT systems were
detected in X-rays. In A0620-00 the quiescent ($\sim$ 2 - 10 keV)
luminosity is $L_X=10^{31}$ erg s$^{-1}$, in the other two systems (GRO
J1655-40 and V404 Cyg) $L_X>10^{32}$ erg s$^{-1}$ (see Garcia et al.
1997 and references therein). This sample is not only small but also
very eclectic as far as companion stars are concerned. The secondary in
A0620-00 is a late type dwarf (K5V, McClintock \& Remillard 1986), in
GRO J1655-40 the F3-6 (Orosz \& Bailyn 1997) secondary is either near
the end of its main-sequence life (Reg\"os, Tout \& Wickramasinghe 1998)
or is crossing the Hertzsprung gap on its way to the giant branch (Kolb
et al. 1997, Kolb 1998), and finally in V404 Cyg the K0 (Casares,
Charles \& Naylor 1992) secondary is a `stripped' giant (King 1993). In
the case of the last system Bildsten \& Rutledge (1999) admitted that
its $L_X= 1.6\times 10^{33}$ erg s$^{-1}$ cannot be emitted by the
companion's corona (they find $L_X/L_{\rm bol}=8\times 10^{-2}$).
However, except for this system and for 4U1543-47 (see below), Bildsten
\& Rutledge (1999) expect quiescent X-ray luminosity of black-hole
LMXTBs to originate in the coronae of secondaries.

I discuss this hypothesis in Sect. 2. and conclude that it cannot be
correct. Black-hole LMXBT's secondary stars cannot be the source of quiescent
X-rays because they are not different from their dwarf nova
counterparts. In the (two known) cases where these secondaries are
different, their coronal X-ray luminosity should be {\sl lower} than in
the corresponding active star binaries, so that also in this case
quiescent X-ray luminosity can only result from accretion. In Section 3
I discuss what the disc instability model of dwarf novae and LMXBTs has
to say about quiescent X-ray emission and in Section 4 I show that, on
simple assumptions, this model combined with an ADAF model (as first
proposed by Narayan, McClintock \& Yi 1996; see also Lasota, Narayan \&
Yi 1996) predicts a correlation between the quiescent X-ray luminosity
and the orbital period. This correlation is satisfied by the three
observed systems, which allows one to make predictions about future
observations by {\sl Chandra} and {\sl XMM-Newton} of systems for which
up to now only upper limits are known. Section 5 ends the article with
discussion and conclusions.
\begin{figure}
\resizebox{\hsize}{!}{\includegraphics{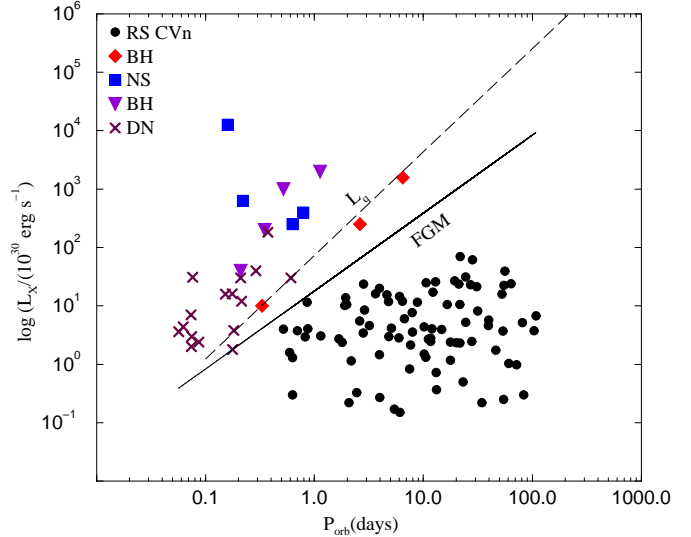}}
\vskip -.3cm
\caption{X-ray luminosities of quiescent dwarf-novae and soft X-ray transients 
and X-ray luminosities of  RS CVn stars. The continuous line marked `FGM' is
the limit given by Eq. (\ref{fgm}) with $M_2=1M_{\odot}$. The dashed line 
corresponds to the relation Eq. (\ref{lx}). Down-pointing triangles correspond 
to upper limits.}
\label{luminosities}
\end{figure}

\section{Companion stars}

Roche-lobe filling, secondary stars in close binaries are rapidly
rotating stars, so their coronae can be rather powerful X-ray emitters by
`normal' star standards, and one could think that they could
explain the quiescent X-ray emission in some close binaries (e.g.
Charles 1996). However, for dwarf novae, and CVs in general, this does
not seem to be possible (Ruci\'nski 1984a; Eracleus et al. 1991; Richards 1996).
Observations of rapidly rotating late type stars suggest a saturation of
X-ray luminosity at approximately (Fleming, Gioa \& Maccacaro 1989)
\begin{equation}
L_{\rm X}\approx 10^{29} \left(R/10^5
{\rm km}\right)^2$~erg~s$^{-1}.
\label{fgmr}
\end{equation}
Using the relation (Paczy\'nski 1971)
\begin{eqnarray}
R_2& = &0.462\left(\frac{M_2}{M_1 + M_2}\right)^{1/3} a\\
a & = & 3.5 \times 10^{10}~\left(\frac{M_1}{M_{\odot}}\right)
^{1/3}~(1+q)^{1/3}~P_{\rm hr}^{2/3}~{\rm
cm},
\label{r2}
\end{eqnarray}
where $M_2$ and $R_2$ are respectively the secondary's mass and radius,
$q$ is the secondary to primary mass-ratio,
$P_{\rm hr}$ the orbital period in hours, one obtains 
\begin{equation}
L_{\rm FGM}=2.7 \times 10^{29} M_2^{2/3}
 ~P_{\rm hr}^{4/3} 
\label{fgm}
\end{equation}
Therefore, coronal X-ray emission from rapidly rotating stars 
has to obey the inequality
\begin{equation}
L_X < L_{\rm FGM}
\label{fgml}
\end{equation}
This observed saturation effect is to be expected on theoretical grounds
(Vilhu 1984, Vilhu \& Walter 1987; Skumanich 1986; Skumanich \& MacGregor 1986). 
Coronal activity and coronal heating are thought to be of
magnetic origin. They depend on the star's rotation speed and on the
depth of the convection layer under the stellar surface. 

As first pointed out by Ruci\'nski (1984a) the relevant measure of
coronal activity is the ratio of the X-ray to bolometric luminosity
$L_X/L_{\rm bol}$. This ratio increases with stellar rotation but gets
saturated at around $10^{-3}$ (see e.g. Singh at al. 1999). However, as
shown both by observation and by models (Fleming et al. 1989), this
saturation is a surface effect: the number of magnetic loops grows until
there is no more space for new ones to appear. As a result the X-ray
luminosity saturates at the limit given by Eq. (\ref{fgmr}). That is why
secondary stars in CVs and LMXBs cannot, despite their fast rotation, be
powerful X-ray emitters: they are just too small (see also Eracleus et
al. 1991).

To illustrate the meaning of the saturation effect let us consider two
cases of rapidly rotating active stars. The very rapidly rotating
($P=9.12$ hr) K2V star called Speedy Mic (HD 197890) has $L_X/L_{\rm
bol} =8.5 \times 10^{-4}$ but its X-ray luminosity is only $8.7 \times
10^{29}$ erg s$^{-1}$ (Singh et al. 1999), in complete agreement with
Eq.~(\ref{fgmr}). The 12.5 hr pre-cataclysmic binary V471 Tau with a K2V
secondaryhas $L_X/L_{\rm bol}> 10^{-3}$ (Ruci\'nski 1984b). Wheatley
(1998) showed that the hard uneclipsed X-ray luminosity may be due to a
coronal activity of the secondary since for this component $L_X/L_{\rm
bol}\approx 10^{-3}$ (the other component results from wind accretion
onto the white dwarf), but its X-ray luminosity is $\sim 10^{30}$ erg
s$^{-1}$ in total agreement with Eq. (\ref{fgmr}). Therefore, if for a
given system $L_X/L_{\rm bol} \approx 10^{-3}$ {\sl and} the X-ray
luminosity is larger than the limit given by Eq.~(\ref{fgmr}) (or
Eq.~(\ref{fgm})), one should rather conclude that this luminosity {\sl
cannot be} due to the coronal activity of the secondary.

The limit given by Eq. (\ref{fgm}) (with $M_2=1M_{\odot}$) is plotted in
Fig. (\ref{luminosities}), which in addition to X-ray luminosities of DN
(Eracleus et al. 1991) and LMXBTs (Garcia et al.~1997 and references
therein) shows X-ray luminosities (Dempsey et al. 1993) of RS CVn stars.
Clearly, X-ray luminosities of all RS CVn stars are below this limit,
whereas X-ray luminosities of dwarf novae (Eracleus et al. 1991) and
LMXBTs (Verbunt 1996) are above it. One can conclude, therefore, that
the X-ray luminosities from quiescent DN and LMXBTs are too large to be
emitted by coronae of stellar companions.

The conclusion about DN was confirmed by observations of eclipsing
systems in which X-rays are clearly emitted near the white dwarf (Wood
et al. 1995; van Teeseling 1997).

Recently, however, Bildsten \& Rutledge (1999) challenged the validity
of this conclusion for black-hole LMXTBs. They estimate the ratio of
X-ray to bolometric luminosity in quiescent black-hole LMXTBs to be
$\sim 10^{-3}$, close to the maximal one observed in RS CVn stars. They
argue, however, that the actual X-ray luminosity of black-hole LMXBT's
secondaries may be larger than in RS CVn because they would be at a
given orbital period, of an earlier spectral type. According to the same
authors the $L_X/L_{\rm bol}$ ratio in both DN and neutron-star LMXTBs
is too high ($\gg 10^{-3}$) for the X-rays to be emitted by stellar
coronae, which confirms previous conclusions.

Even if secondaries in black-hole LMXTBs were of an earlier type than,
say CVs, this would not help much because the saturation value of $L_X$
depends mostly upon radius and not on the effective temperature (Fleming
et al. 1989). In any case, as I show below, if black-hole LMXBT's
secondaries were different from CV's companions at the same orbital
period, they would rather be of a {\sl later} type.

\begin{figure}
\resizebox{\hsize}{!}{\includegraphics{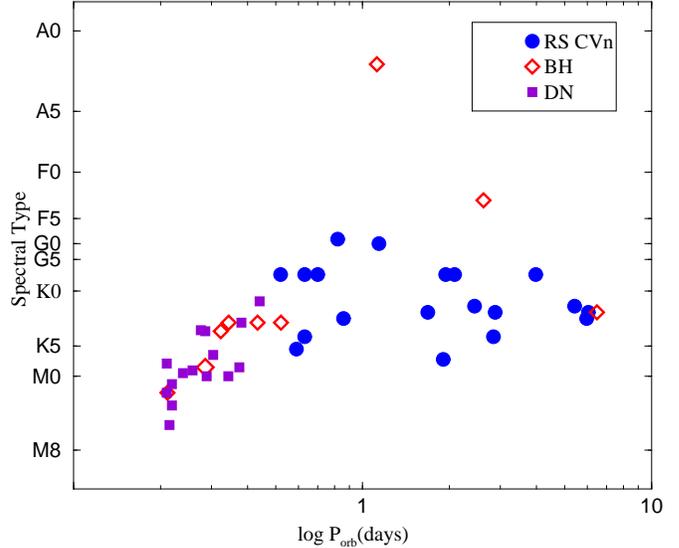}}
\vskip -.3cm
\caption{Spectral types of secondary stars of dwarf novae, black -hole X-ray 
transients and of RS CVn stars }
\label{sptypes}
\end{figure}

Fig. (\ref{sptypes}) shows the spectral types of stellar companions as a
function of the orbital period, for dwarf novae, black-hole LMXTBs and
selected RS CVn stars. Dwarf nova data are taken from Beuermann et al.
(1998) where I selected only systems with orbital periods longer than
$\sim 5$ hr. black-hole LMXTBs spectral types are the same as in
Bildsten \& Rutledge (1999) and the spectral type of GRS1009+45 (X-ray
Nova Velorum 1993) is taken from Filippenko et al. (1999). From
Dempsey's et al. (1993) Table 1, I chose systems of the latest type, in
order to maximize the `chances' of the assertion according to which
secondaries in black-hole LMXTBs are of earlier type. I took the same
attitude towards the error bars or spectral type ranges, choosing the
latest possible for DN and RS CVn's and the earliest for black hole
systems. The statistics is rather poor since only nine black-hole LMXTBs
have known orbital periods. Fig. (2) shows that except for two systems,
4U 1543-47 (A2) and GRO J1655-40 (F3), all the other systems (i.e., the
other six) have spectral types similar to that of DN and RS CVn's at the
same orbital period. Considering that I have chosen the earliest
possible spectral types for LMXBTs and the latest possible for the other
systems, one could conclude that LMXBT secondaries are of {\sl later}
type.

The two odd systems are {\sl less} evolved than secondaries in RS CVn's
and dwarf novae (not plotted here but known to contain subgiants or
giants, see e.g. Warner 1995) at the same orbital period. These two
systems, 4U 1543-47 and GRO J1655-40, are exceptional because their
secondaries are much more massive than in other `low mass' binaries.
They are the only two systems (out of probably 12, see Kalogera 1999 and
references therein) with intermediate mass ($\gta 2 M_{\odot}$)
secondaries. 4U 1543-47 is in phase $A$ of binary evolution, expanding
away from the main sequence (Kolb 1998) and GRO J1655-40 is either close
to the end of its main sequence life (Reg\"os et al. 1998) or is
crossing the Hertzsprung gap (Kolb et al. 1997, Kolb 1998). Late-type
companions of RS CVn's are probably crossing or have already crossed
this gap (Popper \& Ulrich 1977). The A2 secondary in 4U 1543-47 is,
anyway, not a very good candidate for an active star (as also
acknowledged by Bildsten \& Rutledge) since e.g. its Rossby number,
measuring stellar activity, is rather large (in simpler terms: there is
no convective envelope to speak of). The situation of GRO J1655-40 is
not much better since, as shown by Fleming et al. (1989), the saturation
X-ray flux for $B-V<0.6$ steadily decreases with increasing stellar
mass. Therefore, the fact that for this system $L_X/L_{\rm bol} \sim
10^{-3}$ would rather suggest that X-rays {\sl are not} due to stellar
activity. 

Until recently it has been thought that companion stars in CVs are, with
a few exceptions, indistinguishable from main-sequence stars (see e.g.
Warner 1995). Recent, detailed comparison of the observed properties of
CV's secondaries with main sequence (MS) field stars showed that, for
orbital periods $\gta 6$ hours, a considerable fraction of CV's
companions are evolved (Baraffe \& Kolb 1999, 2000; Beuermann et al. 1998;
Beuermann 1999). 
%
At a given orbital period these evolved secondaries
are therefore {\sl cooler} than a hypothetical main sequence star (i.e. of
an {\sl later} spectral type at a given orbital period).
There is no known reason why
secondaries in black-hole LMXTBs should have very different properties (e.g. King
1999), which is confirmed by Fig. 2. Therefore, if coronal emission from
secondaries cannot be the source of X-rays in CVs, the same is true of
black-hole LMXTBs, as suggested in any case by Figs. (1) and (2). Therefore,
X-ray emission from black-hole LMXTBs must be produced directly by accretion.
The same is not necessarily true for neutron-star LMXTBs as proposed by Brown et
al. (1998). I will come back to this problem in Section. 5.

All of this does not mean that secondary stars in LMXBTs are not active
(Ruci\'nski 1984b). They should be active, at least for short period systems
whose evolution is supposed to be driven by magnetic braking. But their
X-ray luminosity is what is expected from such stars: too weak to
explain observations of quiescent LMXBTs.

The quiescent X-ray luminosity may vary by a factor $\gta 3$ in few days
(in Cen X-4, see Campana et al. 2000 and references therein) or by a
factor $\sim 3$ and $\sim 10$ in A0620-00 and V404 Cyg respectively
(McClintock private communication) on a not yet determined timescale.
Such variations are not unexpected in an ADAF, but as pointed out by
Campana et al. (2000) they could also be due to giant flares in coronae
of secondary stars. Such flares are observed in RS CVn's and Algol
systems and their luminosity can reach $\sim 10^{32}$ erg s$^{-1}$
(Ottmann and Schmitt 1996; G\"udel et al. 1999). However, the hope
expressed by Campana et al. (2000), that even more energetic flares
should be expected from shorter period systems, is substantiated neither
by models nor by observations (Audard et al. 2000).

\section{Quiescent X-rays according to the disc instability model}

There is reasonable evidence that both DN and LMXBT outbursts are
triggered by the same physical mechanism: a thermal-viscous instability
due to hydrogen recombination at the disc's effective temperature $5500\
{\rm K}\lta T_{\rm eff} \lta 7500$ K. Any disc in which the effective
temperature is at some point in this instability strip, will be subject
to some kind of outburst. The presence of thermal instability does not
by itself guarantee that outburst properties will agree with
observations. Several ingredients must be added to the disc model in
order to obtain outburst cycles which have anything to do with reality
(e.g. Smak 1984, 1999; Liu, Meyer \& Meyer-Hofmeister 1997; Hameury et
al. 1998, Hameury, Lasota \& Warner 2000; Menou et al. 2000; Osaki
1996). In LMXBTs disc X-ray irradiation is particularly important, as
pointed out by van Paradijs (1996) and elaborated by King \& Ritter
(1998), Dubus, Hameury \& Lasota (2000) (see also Lasota 2000b) and
Esin, Lasota \& Hynes (2000).

According to all versions of this disc instability model the outburst
cycle is due to the disc oscillating between a cold quiescent state and,
in outburst, a hot, fully ionized configuration. In quiescence the whole
disc must be in the cold state. Such a quiescent disc is {\sl not} in
viscous equilibrium (which is the very reason for the outbursts), i.e.
the accretion rate is {\sl not} constant with radius and the matter
transferred by the secondary accumulates in the disc. The presence of a
non-steady disc in quiescent DN is confirmed by observation of eclipsing
systems, which show flat effective temperature profiles (Wood et al.
1986, 1989, 1992; Rutten et al. 1992) as predicted by the model (Smak 1984).

In the DIM the accretion rate in quiescence must satisfy the inequality
(e.g. Hameury et al. 1998)
\begin{eqnarray}
\dot M_{\rm quie}(R)& <& \dot M_{\rm crit}= \nonumber \\
4.0 &\times& 10^{15} \left(\frac{M}{M_{\odot}}\right)^{-0.88}
\left(\frac{R}{10^{10} {\rm cm}}\right)^{2.65} \ {\rm g} \ {\rm s}^{-1}
\label{mdotcr}
\end{eqnarray}
where I omitted a term very weakly dependent on the viscosity parameter $\alpha$.

This means that in quiescence the rate of accretion at the last stable orbit around
a black hole or neutron star primary must satisfy
\begin{equation}
\dot M < \dot M_{\rm quie}
\approx 3.6 \times 10^{4} \left(\frac{M}{M_{\odot}}\right)^{1.77}
  \left(\frac{R}{3R_{\rm G}}\right)^{2.65} \ {\rm g} \ 
{\rm s}^{-1},
\end{equation}
where $R_G=2GM/c^2$ is the Schwarzschild radius. These accretion rates are much
lower than the mass transfer rates from secondaries.

In terms of accretion luminosities this implies the following limit:
\begin{equation}
L_{\rm accr} \lta 6.6 \times 10^{24} 
\left(\frac{\eta}{0.1} \right)\left(\frac{M}{M_{\odot}}\right)^{1.77} 
\left(\frac{R}{3R_{\rm G}}\right)^{2.65} {\rm erg \ s^{-1}}
\label{xtlim}
\end{equation}
for LMXBTs (where $\eta$ is the accretion efficiency).
Fig. (\ref{luminosities}) shows
observed X-ray luminosities from quiescent dwarf novae 
and X-ray transients.
For all LMXBTs 
these luminosities are much higher than the limits given by Eqs.
(\ref{xtlim}). 
(It is sometimes claimed that the quiescent X-ray luminoisties are 
much {\sl lower} than expected. Such claims assume,
that the quiescent disc is stationary which, obviously, is {\sl not}
the case.)
If observed X-rays are emitted by the accretion flow in the vicinity of the
central compact body then the cold quiescent accretion disc whose presence 
is required by the DIM cannot extend down to the central compact body
(Meyer \& Meyer-Hofmeister 1994; Mineshige et al. 1992; Lasota 1996).

The DIM can be reconciled with X-ray observations if the quiescent disc
is truncated at some radius $R_{\rm tr}$ as proposed by Meyer \&
Meyer-Hofmeister (1994) for DN and by Narayan et al. (1996) for LMXTs.
For DN $R_{\rm tr}> R_{\rm WD}$ where $R_{\rm WD}$ is the white dwarf's
radius, for LMXTs $R_{\rm tr}>> R_G$. In the case of Narayan et al.
(1996) the motivation was different: they realized that X-ray, UV and
optical spectra of quiescent black-hole LMXTBs cannot be reproduced by a
standard model (Shakura \& Sunyaev 1973) in which the accretion disc
reaches down to the last stable orbit but are well represented by a
model in which the inner part of the disc is replaced by an ADAF. The
two aspects of the problem, the DIM and the spectra, have been taken
into account by Lasota, Narayan \& Yi (1996), Narayan, Barret \&
McClintock (1997), Hameury et al. (1997) for black-hole LMXTBs and by
Meyer, Meyer-Hofmeister \& Liu (1996) for DN.

During outbursts the inner disc radius reaches the last stable orbit and
then recedes to its quiescent position (Esin et al. 1997; \.Zycki et al.
1998, 1999;, Menou et al. 2000, Dubus et al. 2000). In the main body of
the quiescent disc (not too close to the outer boundary) the accretion
rate radial profile is roughly parallel to the critical rates, i.e. its
radial slope is $\sim 2.6$. Of course, this profile changes and the
accumulating matter creates somewhere a bump which will trigger an
outburst by crossing the $\dot M_{\rm crit}$ line. However, for systems
with very long recurrence times, such as black-hole LMXTBs, one can
consider that during quiescence the accretion rate radial profile is
parallel to the critical one. This results from the roughly self-similar
character of the decay from outburst produced by the cooling wave. As a
result, at a given radius (not too close to the outer boundary) the disc
state always ends up at the same place on the $S$-curve (see Menou,
Hameury \& Stehle 1999b, especially Fig. 9).

One can therefore estimate the accretion rate at the truncation (transition)
radius just by rescaling Eq. (\ref{mdotcr}) to typical parameters and one 
obtains
\begin{equation}
\dot M \approx \dot M_{\rm tr}
\approx 2.4 \times 10^{15} \left(\frac{M}{7 M_{\odot}}\right)^{1.77}
  \left(\frac{R_{tr}}{10^4R_{\rm G}}\right)^{2.65} \ {\rm g} \ 
{\rm s}^{-1}
\label{accr}
\end{equation}
Because of the strong dependence on $R$ this is quite a good estimate of
the actual accretion rate, as I will argue in the next Section. On the
other hand, to make the formula operational one has to decide what (or
where) $R_{tr}$ is.

\section{X-rays from ADAFs in quiescent black-hole LMXTBs}

The main problem of the accretion-disc+ADAF model is the absence of a
satisfactory description of the transition between the two flows.
Although substantial advances have been made in this domain (Kato \&
Nakamura 1999; Liu et al. 1999; Manmoto et al. 2000) the value of the
transition radius is still rather a free parameter than a well
determined physical quantity. Spectral models of quiescent black-hole
LMXTBs suggest a value of the transition radius $R_{\rm tr} \gta 10^4
R_G$ (Narayan et al.1997; Hameury et al. 1997; Quataert \& Narayan
1999). The same result is obtained when one models outbursts of LMXBTs
(Hameury et al. 1997; Menou et al. 2000; Dubus, Hameury \& Lasota 2000).

Menou, Narayan \& Lasota (1999a) found a simple prescription for the
transition radius and used it rather successfully to describe the
stability properties of LMXBTs. They noticed that the ADAF part of the
flow should not extend to radii larger than the impact radius of the
stream of matter that is being transferred from the secondary. The cold
incoming stream would form an annulus at the impact radius $R_{\rm
impact}$. One can expect that this material will spread a little under
the influence of viscous evolution before it evaporates fully. Thus one
expects $R_{\rm tr} \lta R_{\rm impact}$, i.e. a transition radius close
to the ring formed by the transferred matter.

This is equivalent to assuming that in quiescence, the inner disc will
evaporate up to the largest radius allowed by the mass-transfer stream.
I will call this hypothesis `Maximum ADAF Hypothesis' (MAH). The term
`strong ADAF principle' is sometimes used, but here this is not a
principle, but just a hypothesis which, as I show below, can be
tested by observations.

Calculations (Lubow 1989) show that the transferred 
stream may overflow the disc and converge to an impact radius which scales 
as $R_{\rm impact} \simeq 0.48 R_{\rm
circ}$ (this coefficient corresponds to $q<0.5$).  
The circularization radius around the accreting 
body  $R_{\rm circ}$ can be approximated by (Frank et al.
1992):
\begin{equation} 
\frac{R_{\rm circ}}{a} = (1+q)[0.5-0.227 \times \log q]^4.\\ 
\label{eq:rcirc} 
\end{equation} 
Menou et al. (1999a) put therefore 
\begin{equation}
R_{\rm tr}=f_t \ R_{\rm circ} ,\qquad f_t<0.48.
\label{rtr}
\end{equation}
 and showed that observational data 
(widths of $H_{\alpha}$ lines) are consistent 
with $f_t \approx 0.25$ for
most systems, except for V404 Cyg where a value of  $f_t \lta  0.1$ is required.

According to Eqs. (\ref{rtr}), (\ref{eq:rcirc}) and (\ref{r2})) the transition radius 
depends on the orbital period of
the binary system. On the other hand, as explained  in the preceding section, the
accretion rate at the transition radius, where the cold quiescent accretion ends, is
a function of this radius. Therefore, by combining Eq. (\ref{accr}) and Eq. (\ref{rtr})
one obtains the following relation: 
\begin{equation}
\dot M_{\rm ADAF} \approx 1.6 \times 10^{18} f_t^{2.65} P_{\rm day}^{1.77} {\rm g \ s^{-1}}
\label{slope}
\end{equation}
where $\dot M_{\rm ADAF}$ is the rate at which matter enters the ADAF from
the cold quiescent disc. This rate is independent of the primary's mass.
The transition radius $R_{\rm tr}$ is sufficiently distant from the
outer disc's edge for the Eq. (\ref{accr}) to be valid.

Fig. (\ref{luminosities}) shows that the X-ray luminosities of the three
detected quiescent systems satisfy a similar relation 
\begin{equation}
L_X \approx L_q=7.3 \times 10^{31} P_{\rm day}^{1.77} {\rm erg \ s^{-1}}
\label{lx} 
\end{equation} 
Of course, Eq. (\ref{slope}) deals with
accretion rates, whereas Fig. (\ref{luminosities}) represents X-ray
luminosities, so that Eqs. (\ref{slope}) and (\ref{lx}) suggest that the
{\sl X-ray} efficiency is \begin{equation} \eta_X \approx 5 \times
10^{-8} f_t^{-2.65} \label{xeff} \end{equation} For a typical value of
$f_t \sim 0.25$ (Menou et al. 1999a) the X-ray efficiency $\eta_X \sim 2
\times 10^{-6}$. 

One can compare accretion rates (i.e. the X-ray efficiency)
predicted by Eq. (\ref{slope}) with ADAF models for the three black-hole LMXTBs in
which X-ray in quiescence were detected (Narayan, Barret \& McClintock 1997; 
Hameury et al. 1997). This amounts to
determining the values of $f_t$. These models are calibrated by the
X-ray luminosity so this comparison is a good test of the validity of
Eq. (\ref{slope}). 
I compared Eq. (\ref{slope}) with models of
Narayan et al. (1997) and Hameury et al. (1997). 
(In Quataert \& Narayan (1999) accretion rates are slightly lower.)

I obtain the same ADAF accretion rates (and X-ray efficiency)n for very
reasonable values of $f_t$. For A0620-00 I get $0.1 \lta f_t\lta 0.3$,
for V404 Cyg somewhat lower values: $0.06 \lta f_t\lta 0.1$, and for GRO
J1655-40 $f_t \sim 0.12$. As mentioned above, the necessity of a lower
value of $f_t$ for V404 Cyg based on $H_\alpha$ line widths was already
pointed by Menou et al. (1999a) which is rather comforting but not a
proof of the validity of my very simple model. This model, however,
allows one to make predictions about future X-ray observations of known
black-hole LMXTBs which until now escaped detection. The prediction is very
simple: all quiescent X-ray luminosities should be close to the line
$L_q$ in Fig.~(\ref{luminosities}). 

I used W3PIMMS (http//heasarc.gsfc.nasa.gov) to calculate the number of
counts that can be expected to be registered by {\sl Chandra} and {\sl
XMM-Newton}. Results are presented in Table 1, in which I also included
a similar estimate for the three systems detected by the previous X-ray
missions. I consider two power-law spectral models with photon power-law
indices 1.5 and 3.5. Distances and column densities are taken from
Garcia et al. (1997), except when a different reference is given. GRO
J0422+32, GRS1009-45 and GS1124-683 are at the detection limit for both
satellites. A 21.5 ksec observation of GS2000+25 by {\sl Chandra}
(Garcia, McClintock \& Murray 2000) failed to detect a single photon
from this source which is consistent with estimates in Table 1,
especially if the photon index is $\sim 3.5$ as in A0620-00 (Narayan et
al. 1997).

One hopes, of course, that the distance determinations given in Table 1
are reliable.
\begin{table*}
\caption{Predicted X-ray fluxes}
\begin{center}
\begin{tabular}{lrllrcc} \hline \hline
\\
System & $P_{\rm orb}$   & D  &log$N_H$ & Flux & {\sl Chandra} counts & {\sl XMM} counts
\\ 
(1)&(2)&(3)&(4)&(5)&(6)&(7)\\
\\
\hline
\\
GRO J0422+32 &$5.1$ & $2.6^{[1]}$ & 21.3 & 0.6  & $ 25^{(a)} - 15^{(b)}$  & $70^{(a)} - 105^{(b)} $\\
\\
GRS1009-45 & 6.9 & 5$^{[2]}$ & 21.05 & 0.3 & $15^{(a)} - 5^{(b)}$ & $35^{(a)}- 50^{(b)} $\\
\\
GS2000+25 & $8.3$ & $2.7$ & 21.92 &  1.3 &  $25^{(a)} - 5^{(b)}$ &$ 85^{(a)} - 45^{(b)}$  \\
\\
GS1124-683 &$10.4$ & $5$& 21.21 & 0.6  & $25^{(a)} - 15^{(b)} $ &$75^{(a)} - 130^{(b)} $  \\
\\
H1705-250 & $16.8$ & $8.6$& 21.44 &  1.3 &  $45^{(a)} - 20^{(b)}$ & $135^{(a)} - 165^{(b)}$  \\
\\
4U 1543-47  & $27.0 $ & 8 & $21.44^{[3,4]}$ & 1.2 &  $40^{(a)} - 20^{(b)}$ & $125^{(a)} - 150^{(b)}$ \\
\\
\hline
\\
A 0620-00$^{\dagger}$ & 7.8 & 1.2 & 21.29 & 5.8 &$230^{(a)} - 135^{(b)}$ & $620^{(a)} - 1150^{(b)} $\\
\\
GRO J1655-40$^{\clubsuit}$ & 62.9  & 3.2 &21.8 & 20.5 &$ 410^{(a)} - 130^{(b)}$  & $1540^{(a)} - 970^{(b)} $\\
\\
V404 Cyg$^{\spadesuit}$ & 155.3 & 3.5 & 22.4 & 108 & $1240^{(a)} - 130^{(b)}$  & $4400^{(a)} - 1032^{(b)}$ \\
\\
\hline\hline
\end{tabular}
\label{tab:systems}
\end{center}
$^{(a)}$ Photon power-law index = 1.5\\
$^{(b)}$ Photon power-law index = 3.5\\
(1) Only the last three systems were detected by previous instruments\\
(2) Orbital periods in hours (see Menou et al. 1999c and references therein and Filippenko et al 1999).\\
(3) Distances to the systems in kpc (see text).\\ 
(4) Logarithm of the column density \\
(5) Unabsorbed X-ray flux in $10^{-14}$ erg cm$^{-2}$ s$^{-1}$
    from formula Eq. (\ref{lx}).\\
(6) For {\sl Chandra} ACIS-S, integrated over 50 ksec.\\
(7) For {\sl XMM - Newton EPIC PN}, integrated over 50 ksec (per one instrument)\\
$^{[1]}$ Esin et al. (1998)\\
$^{[2]}$ Barret et al. (2000)\\
$^{[3]}$ Orosz et al. (1998)\\
$^{[4]}$ Predehl \& Schmitt (1995)\\
$^{\dagger}$photon-index $\sim 3.5$ from $ROSAT$ observations (Narayan et al. 1997)\\
$^{\clubsuit}$photon-index $\sim 1.5$ from $ASCA$ observations (Hameury et al. 1997)\\
$^{\spadesuit}$photon-index $\sim 2.1$ from $ASCA$ observations (Narayan et al. 1997)\\
\end{table*}

\section{Discussion and conclusions}

The quiescent X-ray luminosity model is rather unsophisticated, but it
is based on several natural properties of the ADAF + DIM-disc model of
LMXTs and has the advantage of making definite predictions that, one
might hope, will be tested by observations. Any detection of GRO
J0422+32, H 1705-25, GS1124-68 would eliminate the coronal model, as
pointed out by Bildsten \& Rutledge (1999). The detection of all five
systems close to the predicted fluxes would count as a success for the
ADAF + DIM-disc model and in particular for the MAH (first suggested by
Narayan \& Yi 1995). One should keep in mind, however, that if the MAH
were to apply in these systems this would not prove its general
validity. In binaries the mass-transfer stream provides a a `barrier'
which prevents the spreading of evaporation. What would form such a
barrier in galactic nuclei is not clear, but in NGC 4258 the ADAF
extends only to few hundred $R_G$ (Gammie, Narayan \& Blandford 1999,
whereas it could in principle extend up to $\sim 10^4R_G $).

The confirmation of correlation between X-ray luminosity and orbital period
given by  Eq. (\ref{lx}) would not necessarily mean that the inner, hot,
accretion flow is an ADAF in the sense of Abramowicz et al. (1995) and
Narayan \& Yi (1995). This correlation means only that the rate at which
accretion enters the hot flow is correlated with the size of the inner
`hole', hence with the orbital period. The `standard' ADAF model is a
good representation of the inner flow. It is not clear that it is only
one. More detailed observational diagnostics is required to decide what
is the real solution.

Quiescent dwarf novae are not expected to satisfy Eqs. (\ref{slope}) or
(\ref{lx}). First, most of them have recurrence times much shorter than
black-hole LMXBs, so that the quiescent accretion rate should vary on a short
time-scales. In fact X-ray quiescent luminosity is highly variable in
these systems (Verbunt, Wheatley \& Mattei 1999). Second, in many cases
the inner disc truncation might be due to the action of the white
dwarf's magnetic field rather than to evaporation (e.g. Lasota et al.
1999). In such a case the transition radius is just the magnetospheric
radius and is independent of the orbital period. 

In the case of neutron-star LMXTBs there are two possibilities. First,
as proposed by Brown et al. (1998), the quiescent X-ray luminosity could
be due to the neutron star surface being heated by thermonuclear
reactions in the matter accreted during outbursts. In this case X-ray
observations would allow the quiescent disc to extend down to the
neutron star (or the last stable orbit). Second, as described in Menou
et al. (1999c), the inner part of the quiescent accretion flow would
form an ADAF as in black-hole LMXTBs. In this case the advected thermal
energy must be emitted from the stellar surface so, for the same
mass-transfer rate, quiescent neutron-star LMXTBs should be more
luminous than those containing black holes, as observed. The problem,
however, is that the model predicts quiescent neutron-star LMXTB
luminosities much higher than observed. In order to reduce the accretion
rate onto the neutron star (by three orders of magnitude) an ADAF model
has, therefore, to involve the `propeller effect' and perhaps the
presence of winds (Menou et al. 1999c), so it is not as simple as the
black-hole LMXBT models (see however Quataert \& Narayan 1999). In this
case, however, as in DN, one would not expect a correlation of the X-ray
luminosity with the orbital period. Observations of variability of
quiescent neutron-star LMXTBs could help in choosing the correct model.
It is worth noting that simulations of outburst cycles of LMXBTs show
that disc truncation is necessary if one wishes to reproduce observed
properties of outburst cycles (Menou et al. 2000; Dubus et al. 2000).

For black-hole LMXTBs things are simpler: all available evidence suggests that
quiescent X-rays are emitted by an ADAF filling the inner part of a
truncated DIM-type accretion disc. A particular version of the model of
such a configuration will be tested by {\sl Chandra} and {\sl XMM-Newton}.

\begin{acknowledgements} 
I am grateful to Mike Garcia for very helpful advice and for information
about {\sl Chandra} observations prior to their publication. Comments
and suggestions by Didier Barret, Guillaume Dubus, Jean-Marie Hameury,
Erik Kuulkers,
Jeff McClintock, Ramesh Narayan, S{\l}awek Ruci\'nski and Nick White
were of great help. I thank Lars Bildsten and Bob Rutledge for
interesting discussions which stimulated me to write this paper. 
\end{acknowledgements}

\end{document}